\documentclass[sn-mathphys,Numbered]{sn-jnl}

\usepackage{graphicx}%
\usepackage{multirow}%
\usepackage{amsmath,amssymb,amsfonts}%
\usepackage{amsthm}%
\usepackage{mathrsfs}%
\usepackage[title]{appendix}%
\usepackage{xcolor}%
\usepackage{textcomp}%
\usepackage{manyfoot}%
\usepackage{booktabs}%
\usepackage{algorithm}%
\usepackage{algorithmicx}%
\usepackage{algpseudocode}%
\usepackage{listings}%
\usepackage{enumitem}

\begin{document}

\title[MOND as a transformation between non-inertial reference frames via Sciama's interpretation of Mach's Principle]{MOND as a transformation between non-inertial reference frames via Sciama's interpretation of Mach's Principle}

\author[1]{\fnm{Manuel} \sur{Uruena Palomo}}\email{muruena7@alumno.uned.es}\email{muruep00@gmail.com}\email{independentphysicsresearch@gmail.com}\date{September, 2024}

\affil[1]{\orgdiv{Fundamental Physics}, \orgname{National University of Distance Education UNED}, \orgaddress{\street{PS/ Senda del Rey nº 11}, \city{Madrid}, \postcode{28040}, \country{Spain}, \text{ORCID: 0000-0002-2327-9193} \text{Independent Physics}}}

\abstract{Moderhai Milgrom's Modified Newtonian Dynamics (MOND) correction to Newtonian gravity or inertia is shown to be equivalent to a more fundamental formulation considering a non-inertial local reference frame and the fixed background of the observable universe, in the spirit of Mach's principle. Both Newton's gravitational constant $G\sim c^2/(M_u/R_u)$ and Milgrom's MOND acceleration scale constant $a_0\sim GM_u/R_u^2$ are replaced by two varying, measurable, and cosmological quantities determined by the causally connected mass and size of the universe. They arise from an inverse and an inverse squared distance scalar fields of matter density, respectively. This Machian interpretation of MOND is invariant under global rescalings of mass, length, and time across all regimes and is free from fundamental constants and free parameters, except for the speed of light. Machian MOND satisfies the fundamental consequences of Mach's principle not featured in Newton's and Einstein's theories: the decrease of inertia of a body when masses are removed from its neighborhood, and in the absence of a cosmic background, rotational motion is undefined up to the speed of light. Consequently, Machian MOND provides the necessary limiting behavior to which any phenomenological non-linear theory of modified inertia or gravity that incorporates Mach’s principle, in agreement with galaxy rotation curves, should reduce as an effective approximation.}


\maketitle

\text{\textbf{Date:} January, 2026}

\newpage

\section{Introduction}\label{sec1}

\subsection{Modified Newtonian Dynamics MOND}\label{subsec1}

The solution to the problem of the perihelion precession of Mercury first involved various failed attempts proposing the unobserved existence of a new small planet Vulcan, an asteroid belt, or a gas cloud orbiting the Sun. However, the resolution came as a modification to the laws of gravity by General Relativity (GR). Similarly, the observation of rotational velocities in disks of galaxies higher than predicted by Newtonian gravity, as depicted in galaxy rotation curves, has driven further research into modifications to the laws of gravity and mechanics. The discrepancies between observed luminance to gravitational mass ratio in galaxies and their missing Keplerian falloff in their velocity curves, together with faster radial velocities of galaxies in galaxy clusters than Newtonian predictions, among other evidence, are referred to as the ‘dark matter' problem \cite{Zwicky1933,Rubin1983,Review}. Modifying gravity or inertia instead of proposing the existence of physical cold dark matter is motivated by the Tully-Fisher (TF) law (the asymptotic velocities in rotation curves are proportional to the fourth root of their baryonic mass), Renzo’s rule (features in visible mass strongly correlate with features in velocity profiles), and the cuspy halo problem (Newtonian gravity generally predicts velocity profiles in galactic cores with visible mass only). These recent modifications to the classical law of gravity or laws of mechanics are called Modified Newtonian Dynamics (MOND) models (or MilgrOmiaN Dynamics), with the first and main MOND being Milgrom’s MOND \cite{Milgrom1983a,Milgrom1983b,Milgrom1983c}, which can be thought of as a modification to inertia (applied to Newton's second law of motion) or to the inverse square law of gravity. Almost all full-fledged versions of MOND are just modified gravity, in which the inverse-square law of gravity approximately changes to an inverse law, and not modified inertia. As modified inertia, MOND modifies the kinetic term in the particle Lagrangian $mv^2/2$ or the ‘inertia term' $mg$, and measured kinematic accelerations are not simply proportional to forces per unit mass.

The reason for the anomalous motion of Mercury was that Newtonian gravity is not a good enough approximation for strong gravitational fields, such as near the Sun, where non-linear GR effects become significant. Milgrom's MOND exceptionally explains the observed rotation curves of galaxies by imposing a correction based on two observational relationships: the TF law, and that the mass or velocity discrepancies (between observed and Newtonianly estimated) are always observed below a particular acceleration scale. The simplest correction that satisfies these observational conditions results in an asymptotic rotational velocity $v_{\infty}=\sqrt[4]{GMa_0}$ independent of distance, with the need to introduce a constant of acceleration $a_0$. The connection between the classical Newtonian regime where $v=\sqrt{GM/r}$ and Milgrom's deep MOND or low acceleration regime is determined by an ‘interpolating function' $\mu \left(g/a_0\right) = 1/ (1+ \left(a_0/g\right)^n)^{(1/n)}$ (with $n=1$ for the simple and $n=2$ for the standard interpolating functions) based on $g$ as the ‘true acceleration' and a constant of acceleration $a_0$ (also referred to as the Hubble acceleration or the acceleration scale constant), which is a free parameter adjusted by data fitting and is in principle considered a fundamental and universal constant in Milgrom's MOND. At the deep MOND regime, $g=\sqrt{g_Na_0}$ with $g_N$ the Newtonian gravitational acceleration, and the effective potential is logarithmic at large distances $V(r)\sim\sqrt{GMa_0}\ln(r)$, diverging at infinity. Milgrom's MOND is only length and time scale-invariant when $a_0\rightarrow \infty$ fixing $Ga_0$ to achieve constant velocities independent of radius at the flat part of the galaxy rotation curves. The interpolating function must satisfy the conditions $\mu(x) \rightarrow 1$ for $x\gg 1$ to recover Newtonian laws and $\mu(x) \rightarrow x$ for $x\ll 1$ to satisfy the TF law. A fixed-distance or fixed-mass (independent of the system under study) modification alone cannot account for the observed rotation velocities due to the TF law, i.e., it is not possible to simply set a constant of distance or a constant of mass to modify the laws of gravity to fit all rotation curves of galaxies (because some galaxies exhibit the effect of dark matter at small radius while others at large radius, and the same happens with different galaxy masses). However, combining both mass and distance in acceleration or gravitational field intensity, Milgrom's MOND achieves this limit where the effect of dark matter in galaxies is generally observed. Still, Milgrom's MOND, being insufficient to explain all dark matter related phenomena such as galaxy cluster dynamics, \cite{Sanders1998,Sanders2003}, being non-relativistic, and not explaining phenomenologically the origin of the interpolating function, is an effective theory or approximation to a more fundamental modification to gravity or inertia.

The acceleration constant in Milgrom's MOND $a_0 \sim 10^{-10} \ m/s^2$, was already related to the cosmological scale through the Hubble constant $H_0$ (or the cosmological constant $\Lambda$) $a_0 \sim cH_0 \sim c^2 \sqrt{\Lambda} \sim c^2/R_u$ originally by Milgrom \cite{Milgrom1983a}, stating that this coincidence could point to a basic theory underlying MOND's phenomenology. In particular, he stated that “\textit{an attractive possibility is that MOND results as a non-relativistic, small-scale expression of a fundamental theory by which inertia is a vestige of the interaction of a body with ‘the rest of the Universe', in the spirit of Mach’s principle.}", “\textit{It seems to me that in looking for an ultimate theory, the Mach principle may serve as a most useful guide}", “\textit{The possible
connection between such numerical “coincidences,” which relate parameters of “local” physics to cosmological parameters and the Mach principle, is quite obvious}", and “\textit{We thus envisage inertia as resulting from the interaction of the accelerated body with some agent field, perhaps having to do with the vacuum fields, perhaps with an “inertia field” whose source is matter in the “rest of the universe”–in the spirit of Mach’s principle}" \cite{Milgrom1993,Milgrom1983a}, although he did not further develop MOND in relation to Machian ideas.

Progress in fundamental physics has been made by reducing the number of fundamental constants (such as the gravitational acceleration of the Earth and other planets for a universal gravitational constant, the unification of the speed of light with the vacuum permittivity and permeability, and atomic constants for the Planck constant). Therefore, it is of considerable interest to explore modifications of the laws of gravity or mechanics by removing fundamental constants.

Various theories of modified gravity or inertia are based on Milgrom's MOND acceleration constant: John Moffat's MOG scalar–tensor–vector gravity approach to galaxy rotation curves is based on two parameters, which are set to fit the galactic Milgrom's MOND's acceleration constant \cite{Brownstein_2006}, Erik's Verlinde's entropic gravity attempts to propose an underlying framework for Milgrom's MOND's acceleration constant \cite{Verlinde2017}, and Deur's self-interacting gravitons model correction to Newtonian gravity contains a physical constant counterpart to Milgrom's MOND's acceleration scale constant \cite{Deur2003}. Others, such as quantized inertia, rely on the acceleration parameter $a_0=c^2/R_u$ where $R_u$ is the radius of the cosmological comoving horizon \cite{McCulloch2007}.

\subsection{Mach's Principle and modified inertia}\label{subsec2}

Newtonian gravity and classical mechanics are founded on the principles of absolute space, absolute time, and absolute motion (including acceleration). These were originally opposed by Leibniz (and others, such as Berkeley in his 1721 \textit{De Motu}, and Huygens) arguing that as observers, we can only epistemically access relative notions of space, time, and motion. Newton's justification of absolute space and motion is portraited in his rotating bucket of water experiment at the beginning of his 1687 masterpiece \textit{Philosophiæ naturalis principia mathematica}, which can be summarized as follows: a bucket with water is set spinning around its axis and, at first, the walls of the bucket rotate relative to the stationary water while the surface of the water remains flat as prior to the spinning. After the water starts to rotate as well, it rises towards the walls and its shape is no longer flat when the spinning bucket and the water are at rest relative to each other. Interactions between water and walls (due to their relative velocities) cannot explain the shape of the water, and Newton assumed that this experiment could be used to measure rotation with respect to an absolute space, “\textit{without relation to anything external, which remains always similar and immovable}". 

The definition of absolute space, time, and motion suffer from circular reasoning. In Newtonian mechanics, absolute space is defined as the frame absent of inertial forces (also referred to as fictitious forces, which come from our choice of reference frame), but the absence of inertial forces is used to prove and identify the existence of absolute space in Newton's rotating bucket of water experiment. According to Newton's first law of motion, an object moves inertially if it is free from outside influences, but the fact that it is free from outside influences is inferred only by observing that it moves inertially. How can the inertial system be unequivocally identified? In Special Relativity (SR), light is measured to travel at the same speed in all inertial frames, and absolute acceleration is defined relative to inertial frames. But inertial frames are defined as frames absent of acceleration, and accelerometers always measure acceleration with respect to a reference frame of calibration that must be assumed to be inertial in the first place. 

Leibniz's relative space, time and motion ideas were further developed by Ernst Mach in his 1883 \textit{The Science of Mechanics}, in which he criticized Newton's conclusion of his bucket of water experiment stating that “\textit{No one is competent to say how the experiment would turn out if the sides of the vessel increased in thickness and mass till they were ultimately several leagues thick.}”\cite{Mach1881}. According to Mach, the relative rotation of the water with respect to the bucket produces no noticeable centrifugal forces, and such forces are instead produced by its relative rotation with respect to Earth and the other celestial bodies (in this way, the relative motion between the several leagues thick bucket and the water could produce noticeable centrifugal forces in the water, and perhaps their non-relative motion could reduce these forces when both rotate with respect to the rest of the universe). In contrast to Newton, who attempted to explain the physical effects of inertia through a sort of resistance to motion within an unobservable absolute space with no physical properties, Mach conceived inertia as an interaction that required other external bodies to manifest. Inertia would depend upon an interaction between masses. Newton did not suspect that the change in the water's shape could be due to rotation relative to the rest of the universe because his action at a distance gravitational force from an infinite and homogeneous universe acting on the water cancels out according to Newton's shell theorem and is independent of velocity or acceleration. But no local phenomena can ever be isolated from the rest of the universe, and its effect certainly reaches the water even if the sum of these forces is zero. According to Mach, and in opposition to Newton's conclusions, a spinning bucket of water in an empty universe would not change the water's shape (one could not detect relative motion in an empty universe, it would be undefinable), and if the rest of the universe was set spinning while the bucket was at rest, the water surface would curve, as both the spinning of the bucket and water or the spinning of the universe are indistinguishable systems without an absolute space. Thus, for Mach, the relativity symmetry of acceleration is broken by the existence of a background matter universe.

Moreover, there is an exact coincidence between the local measurement of the angular velocity of the Earth by Foucault's pendulum and the cosmological measurement through the apparent movement of distant stars and galaxies, which Newtonian gravity and mechanics cannot explain because it does not causally connect both measurements (and consider the determination of inertial frames by the fixed stars and Foucault's pendulum a coincidence), but constitutes the basic idea of Mach's principle: “\textit{The universe, as represented by the average motion of distant galaxies, does not appear to rotate relative to local inertial frames.}”\cite{Bondi1997}. For Mach, its the distant masses the ones which hold the pendulum fixed with respect to the Earth.

According to Mach, and known as Mach's principle, the inertia of a body is not an independent and intrinsic property of matter (unlike in Newtonian mechanics, where a particle in an empty universe has the usual inertial properties), but rather the result of the action of the universe as a whole. Mach suggested that the fixed background distribution of matter in the universe must exert the inertial forces on a local accelerating body. In this way, the reference system with respect to which the universe is at rest or in uniform and rectilinear motion is a true inertial reference system. In Ernst Mach words, “\textit{I have remained to the present day the only one who insists upon referring the law of inertia to the Earth, and in the case of motions of great spatial and temporal extent to the fixed stars.}" \cite{Mach1881}. Hence, inertial frames should be defined with respect to this rest frame, and local physical laws must be determined by the large-scale structure of the universe.

Einstein's pursuit of a relativistic gravitational theory was inspired by Mach, and he envisioned GR to fulfill his interpretation of Mach's principle: “\textit{the metric tensor is completely caused and determined by the stress-energy tensor}" \cite{Einstein1918}. He noted that following Mach, it was expected that “\textit{1. The inertia of a body must increase when ponderable masses are piled up in its neighborhood, 2. A body must experience an accelerating force when neighboring masses are accelerated, and, in fact, the force must be in the same direction as the acceleration, 3. A rotating hollow body must generate inside of itself a “Coriolis field”, which deflects moving bodies in the sense of the rotation, and a radial centrifugal field as well}" \cite{Einstein1922}. He had hopes that by using tensors in GR, he would achieve the strong Machian version of relativity (in which the gravitational influence of the whole universe gave rise to inertia). But although consequences 2. and 3. were satisfied in GR (at least phenomenologically by frame-dragging), the first condition was not, because the immediate consequence of it is that a body in an otherwise empty universe should have no inertia, and the empty solution where the stress-energy tensor is zero everywhere in GR is the flat Minkowski spacetime of SR, in which test bodies have the usual inertia. He also referred to this consequence in 1913 stating that “\textit{It must be demanded that the inertial resistance of a body could be increased by having unaccelerated inertial masses arranged in its vicinity; and this increase of the inertial resistance must disappear again if these masses accelerate along with the body}". GR can be considered more Machian, since the metric defining inertia is dynamical and can be acted upon (in contrast to the case of Newton's or SR, where the metric cannot be affected, and thus, its an absolute). Einstein then introduced the cosmological constant term (as the mass density of the universe) in his field equations in the hope that, with the cosmological term, they would have no solutions for a zero stress-energy tensor. However, de Sitter found in 1917 a solution for the field equations with the cosmological constant and with zero stress-energy tensor for an expanding universe, Einstein dismissed the cosmological constant as no longer justified, and he abandoned the ideas of Mach. In his effort to render GR fully Machian, Einstein reached the conclusion that the metric should diverge at infinity (in particular, that the time component of the metric tensor, associated with the Newtonian gravitational potential, should should increase without bound at infinity, and the spatial component should vanish) to satisfy the relativity of inertia (Mach's principle): “\textit{It seems, therefore, that such a degeneration of the coefficients \textnormal{[of the components of the metric tensor]}, is required by the postulate of relativity of all inertia. This requirement implies that the potential energy $m\sqrt{B}$ \textnormal{[with $ds^{2} = -A d\vec{x}^{2} + B dt^{2}$]} becomes infinitely great at infinity.}"\cite{Einstein1917}. Unfortunately, Einstein thought at the time that the stars at infinity were moving very slowly, making the divergence of the metric incompatible with this belief, and dismissed the divergent metric for the cosmological constant due to his lack of knowledge of the expansion of the universe.

Einstein's equivalence principle (a frame linearly accelerated relative to an inertial frame in the Minkowskian spacetime of SR is locally identical to a frame at rest in a uniform gravitational field) axiomatically assumes that inertial and gravitational masses are equal. Gravity is the only infinite range force that cannot be screened due to the existence of only positive masses, unlike in electromagnetism, where positive and negative charges result in balance and neutrality, and astrophysical and cosmological structures are therefore not governed by electromagnetism. Moreover, there always exists a reference frame in which inertial forces vanish, just as for the case of gravitational forces in free fall, and the gravitational field, unlike the electromagnetic field, carries the information about a particle's rest mass. These equivalences indicate that both inertial and gravitational forces are of the same nature. Following Mach, in any two-body interaction, the influence of all other matter inside their causal spacetime should be taken into account. Inertia is thus a form of gravitational induction, appearing when a body is accelerated with respect to the rest of the universe. But Newtonian gravity and GR were developed without considering today's knowledge of the mass content and size of the observable universe. Then, it seems that to describe, for example, the Earth's motion around the Sun, only local masses and distances are required, but the Machian perspective implies that the universe's action is already taken into account in the Newtonian laws. The only way this can be true is through the Newtonian gravitational constant, which, together with the postulate of the inertial frame of absolute space, both constitute the two arbitrary choices of Newton.

In Newtonian gravity, the gravitational constant $G$ is the only fundamental constant of the theory, and it is known with far less precision than any other fundamental dimensional constant in physics. The uncertainty in its measurement, together with its problematic in Quantum Field Theory with the Planck energy (which is used as a cutoff for the energy density of the vacuum and the theoretical quantum corrections estimation of the Higg's mass, leading to the vacuum catastrophe and hierarchy problems) has led to questioning its constant nature repeatedly over the last century; see \cite{Uzan2011} for constraints. GR carries this constant (a varying $G$ would imply a violation of the strong equivalence principle, but only the weak principle is supported by the very precise Eötvös experiments) with the addition of the speed of light, which can be derived from the vacuum permittivity and permeability constants. Therefore, it is reasonable to question whether the gravitational constant is also a derived parameter.

The first historically suggestion of gravity arising necessarily as a consequence of the relativity of inertia came from Hans Reissner \cite{Reissner1915}. Schrödinger soon identified in 1925 \cite{Schrodinger1925} that GR did not fully implement Mach's principle, and proposed a relationship between the gravitational potential of distant masses and the speed of light: “\textit{This remarkable relationship states that the (negative) potential of all masses at the point of observation, calculated with the gravitational constant valid at that observation point, must be equal to half the square of the speed of light.}". This conclusion was reached by imposing that the kinetic energy term had an origin in a potential-like interaction (such as all other forms of energy, which have an origin in an interaction) and was relational, so that $K=mv^2/2$ can be expressed through the gravitational Newtonian potential $Gm/r$ so that $K=(Gm_{i}/ r_{ij})(m_{j} v_{ij}^2 /2c^2)$ with the need to introduce the square of the speed of light. The term $Gm_{i}/r_{ij}c^2=GM_u/c^2R_u \sim 1$ is considered missing in the original Newtonian formulation, with $M_u$ being the sum of all masses within $R_u$, the distance between the moving particle $m_{j}$ and $M_u$. Schrödinger's formulation predicted a finite observable universe, since the potential of an infinite non-expanding eternal universe would also be infinite, even though the expansion of the universe was not known at the time. This relationship was later credited to Reissner's 1915 work by Schrödinger himself.

Schrödinger's formulation appeared repeatedly in several works after him \cite{Whitrow1946,Treder1972,WhitrowRandall1951,Dicke1961,Assis1999}, but it was popularized by Dennis Sciama, who developed in 1953 a modified cosmological vector potential theory of inertia in which the inertial law arises as a side effect of gravity \cite{Sciama1953}. Sciama realized that in the analogous case of Maxwell's equations applied to gravity, the rate of change of the vector potential leads to a term in the ‘gravitoelectric' field that depends on the acceleration of an object relative to the rest of the masses of the universe. He showed that local inertial forces result from the gravitational induction of the universe, so that the dynamics of a rotating body would be affected by the large-scale distribution of the mass of the universe. Sciama postulated that “\textit{In the rest-frame of any body the total gravitational field at the body arising from all the other matter in the universe is zero}”. The induced inertia decreased with $1/r$ with $r$ being the distance to the inertial body, so that the action of global matter dominates over the action of local matter, and it is not significantly modified by the acceleration with respect to local matter, leading to the illusion that inertia depends only on the body itself. The gravitational constant at any point was determined by the total potential of the distribution of matter of the universe, taking the form of $G\sim c^2/(\Phi_u + \varphi)$ with $\Phi_u=M_u/R_u$ and local $\varphi=M/r$ (assuming a linear superposition) which reduces to $G=c^2 /( M_u/R_u)$ in most cases (when locally $\Phi_u>>\varphi$, since only near neutron stars or black holes $\varphi \sim \Phi_u$) with $M_u$ and $R_u$ the mass and radius of the observable universe, causally connected by the speed of light in the past light-cone of the point considered to evaluate $G$, for the equivalence of inertial and gravitational masses to hold. In this way, local phenomena were strongly coupled to global properties of the universe. Sciama knew about the Hubble expansion of the universe, but his model predicted an almost flat observable universe in which critical cosmic matter density is reached for $G\Phi_u/c^2 =1$ to hold, even though the flatness of the universe was not observed at his time.

By pure dimensional analysis, the Reissner-Schrödinger-Sciama's relationship (also known as the Hofmann-Reissner-Schrödinger relationship, henceforth Sciama's relationship) is the only possible derivation of the units of measurement of the gravitational constant through a mass, a distance, and the speed of light. If Sciama's relationship is derived from a more fundamental theory, it can be expected to also contain integers and mathematical constants, but these will be omitted for simplicity, since it does not significantly affect the resulting orders of magnitude of the comparison between $M_u$ and $R_u$. From this relationship, it became evident that the inability to explain the gravitational constant in both Newton's and Einstein's theories is related to the inability to explain inertia à la Mach and to consider the universe at large.

Carl H. Brans introduced a thought experiment considering the ratio between the inertial mass of a body and the active gravitational mass following Mach's principle in a static universe consisting only of a mass shell of radius $R_u$ and inertial mass $M_u$ (these will be the mass and size of the universe), together with a relatively small body of inertial mass $M$ and a test particle. He expected the acceleration of the test particle to be independent from the mass of the particle itself (from the equivalence principle), to depend on the mass $M$, and because of Machian arguments about the inertial properties of the test body being determined by the distribution of matter in the universe and not intrinsic of the body, conceivably to also depend on $M_u$ and $R_u$. Thus, $a=f(M_u,R_u)M/r^2$ and he related $f$ to the gravitational constant \cite{Brans1962}. Together with Robert Dicke, they attempted to introduce Sciama's relationship $G=c^2/(M_u/R_u)$ in a relativistic scalar-tensor theory of gravitation in 1961 \cite{Dicke1961}, inspired by Mach's principle, by interpreting the scalar field as an advanced wave integral over all matter. One interpretation of Brans-Dicke theory consists in the variation of the inertial mass with the scalar field if the gravitational constant is constant, so that Einstein's field equations are formally valid, but the most well-known interpretation allows the gravitational constant to be variable with position and time. The difference in predictions between the Brans-Dicke theory and GR has been locally tested in the solar system near the Sun through the Shapiro time delay effect measured by the Cassini experiment, showing that the Brans-Dicke dimensionless coupling factor of the scalar field to gravity, which differentiates between the theory and GR, must be so high that Brans-Dicke theory is almost indistinguishable from GR. All of these tests have been performed in the high-acceleration regime of the solar system.

Hans-Jürgen Treder developed a model similar to that of Schrödinger in a more complete inertia-free mechanics model \cite{Treder1972} based on the Riemann potential as a velocity-dependent gravitational potential (instead of the Weber potential of Reissner and Schrödinger). This model implements the idea of inertia having a gravitational origin without Schrödinger's anisotropic inertial mass (which is ruled out by observations, such as the Hughes–Drever experiments) with Sciama's $G\sim c^2/(\Phi_u + \varphi)$, in agreement with Mach's principle. It is worth noting that Treder was able to implement Mach's principle maintaining a scalar inertial mass instead of a tensorial one, exhibiting no preferred direction and thus avoiding inertial anisotropy. In this non-relativistic linear model, the weak equivalence principle can be derived and not axiomatically postulated as in GR, the weakness of the gravitational constant is explained, and inertial mass can also derived from the model. 

Sciama's relationship with today's measurements of the current cosmological model is satisfied only in terms of orders of magnitude, because the Hubble tension impedes a precise calculation. It yields the orders of magnitude of the observed gravitational constant when considering not the baryonic and dark matter content for $M_u$, but the total energy content (since all forms of energy must gravitate according to GR), which roughly corresponds to the critical density of the universe as measured to be almost flat. As Milgrom stated in one of his conferences, “\textit{the only system that is strongly general relativistic and in the MOND regime is the Universe at large}". However, the critical density is calculated with the assumption of a \textit{constant} gravitational constant itself and considering dark matter, which would not contribute if a correction to the laws of gravity and inertia accounts for its effects, as it will be proposed in the next section.

Few authors have attempted to develop a MOND for galaxy rotation curves based on Mach's principle \cite{Unzicker2003}, or relate Milgrom's MOND to Sciama's relationship and Mach's principle \cite{Darabi2010,Corda2016,Anderson2011,Gine2009,Gine2012,Abreu2014,Corda2024}. The most relevant ideas in the literature relating these topics are summarized onwards.

Alexander Unzicker explored galaxy dynamics in relation to Mach's Principle and the rotating bucket of water experiment, and proposed a basis for a MOND to solve flat rotation curves without considering Milgrom's MOND \cite{Unzicker2003}. He draws attention to the following thought experiment. According to Mach, the system of two distant masses rotating around their center of mass in equilibrium of gravitational and centrifugal forces in an empty universe, since rotation between them is undefined due to absence of absolute space, must be equivalent to the system of those two distant masses not rotating and without gravity (only radial or relative velocities are measurable, and tangential velocities are impossible to measure instantaneously, as they are indistinguishable from a rotation of the coordinate system of the observer). The maximum possible angular velocity is limited by $w_{max}=c/r$ so that the maximum tangential speed is $v=c$ for one of the masses rotating around the other one, since the speed of light cannot be surpassed in any case. This luminal rotational speed in absence of a background universe from Unzicker's thought experiment is a feature of introducing Sciama's relationship in Newton's law of gravity, in which the Newtonian $v=\sqrt{GM/r}$ changes to $v=c\sqrt{(M/r) / (M_u/R_u)}$ after introducing $G=c^2/(M_u/R_u)$ and without a background universe $M_u \rightarrow M$ and $R_u \rightarrow r$, velocities are $v=c$. In other words, Sciama's relationship has the equivalent effect of decreasing inertia when decreasing $M_u$.

Jaume Gine explored a Machian interpretation of the modified second law of motion for the simple interpolating function of Milgrom's MOND based on the accelerated expansion of the universe \cite{Gine2009}, in which $a_0=GM_u/{R_u}^2=c^2/R_u$ is the acceleration that an experimental body feels induced by the rest of the matter of the universe in its inertial frame of reference, and the value of acceleration for which inertial and gravitational masses of a body can differ. Thus, the equivalence principle between inertial and gravitational masses is broken for accelerations smaller than $a_0$. In a following paper \cite{Gine2012}, Gine attempted to derive a phenomenological version of Milgrom's MOND modification to Newton's second law. He considers $a_0$ to be the acceleration at which the edge of the universe at distance $R_u$ goes away from a considered central inertial point due to the expansion of the universe. By relativizing acceleration considering the distant universe at rest, he derives a new interpolating function similar to Milgrom's MOND simple interpolating function.

A complete FundaMOND theory remains only a faint hope until the origin of Milgrom’s MOND is correctly identified. Even though interesting ideas have been put forward relating MOND to Mach's principle, a formulation based on Einstein's Machian consequence not satisfied by GR, by which the inertia of a body should decrease when masses are removed from its neighborhood, will be presented in the next section. Such formulation should arise as an approximation in any non-linear Machian theory of modified inertia or gravity which attempts to explain galaxy rotation curves. The motivation behind this interpretation of Milgrom's MOND is that MOND can be reformulated with the same parameters of the mass and radius of the observable and causally connected universe that the Machian models of modified inertia are built on.

\newpage

\section{Machian MOND}\label{sec2}

Starting from the well-known form of Milgrom's MOND standard \textit{or} simple interpolating function correction to Newtonian gravity \textit{or} to inertia, the ‘true' gravitational acceleration $g$ \textit{or} ‘true' second law of motion is

\begin{equation}\label{eq1}
    g = \frac{G}{\mu \left(\frac{g}{a_0}\right)} \frac{M}{r^2}=\frac{g_N}{\mu \left(\frac{g}{a_0}\right)} \ \ or \ \ F=\mu \left(\frac{g}{a_0}\right)m_ig
\end{equation}

\begin{equation}\label{eq2}
     \mu\left(\frac{a}{a_0}\right) = \frac{1}{\sqrt{1+\left(\frac{a_0}{a}\right)^2}} \ \ or \ \ \mu\left(\frac{a}{a_0}\right) = \frac{1}{1+\frac{a_0}{a}}
\end{equation}

\noindent with $a_0$ being Milgrom's MOND acceleration scale constant, $g_N=GM/r^2$ the Newtonian gravitational acceleration, $M$ the active gravitational mass, $m_i$ the inertial mass, and $r$ the distance between the centers of mass of the active and passive gravitational masses. For the motion of a test particle in a gravitational field in the low acceleration or deep MOND regime, $\mu\left(a/a_0\right) \rightarrow a/a_0$ and $g=\sqrt{a_0 g_N}$, so that with $g=v^2/r$, $v=\sqrt[4]{GMa_0}$ to achieve flat galaxy rotation curves in agreement with the TF law in a remarkable simple way with a single new constant $a_0$. For the high acceleration regime, $\mu\left(a/a_0\right) \rightarrow 1$ and Newtonian laws are recovered. One can also define an equivalent interpolating function $g=\nu(a_0/a_N)a_N$ converting the Newtonian gravitational field into a Milgromian field, with $\nu(a_0/a_N)=(1/2)^{1/n}(1+\sqrt{1+4(a_0/a_N)^n})^{1/n}$ (with $n=1$ for the simple and $n=2$ for the standard interpolating functions).

Milgrom's MOND acceleration scale constant $a_0$ is usually interpreted as a fundamental constant in MOND, but possibly related to the Hubble parameter $a_0\sim cH_0$ or the cosmological constant $a_0\sim c^2\sqrt\Lambda \sim c^2/R_u$ as the energy density of the vacuum $\rho_{vac}$ with $\Lambda=8\pi G \rho_{vac}/c^2$ according to GR, which coincides in orders of magnitude (a better match can be done with $a_0=c^2\sqrt{\Lambda/3}$, although integers and Pi are onwards omitted for simplicity). This is due to Milgrom's wish to define absolute acceleration with respect to a smooth quantum vacuum energy frame in MOND as modified inertia. However, the cosmological constant is only measured indirectly, and the connection between dark energy and a non-zero energy density of the quantum vacuum remains, at best, an optimistic hypothesis. In contrast, Mach's principle implies that acceleration should be defined with respect to the mass distribution of the universe, and the mass and size of the universe are based on direct, albeit approximate measurements. 

In order to express $a_0$ in terms of the mass and radius of the universe, the cosmic coincidence implies $\rho_{vac} \sim \rho_{u}\sim M_u/{R_u}^3$ with $M_u$ the causally connected mass to the local system of study through the radius $R_u$ of the observable universe (or particle horizon radius). Newton's gravitational constant $G$ is substituted for Sciama's interpretation of Mach's principle $G\sim c^2/(M_u/ R_u)$, so that $a_0\sim GM_u/{R_u}^2$, and by reverse engineering, $v\sim c\sqrt[4]{M/M_u}$ and $g\sim c^2\sqrt{M/M_u}/r$ in the equivalent deep MOND regime. By pure dimensional analysis, $a_0\sim GM_u/{R_u}^2$ is also the only possible equation to obtain an acceleration from $M_u$, $R_u$ and other fundamental constants. The interpolating function is now invariant under global rescaling of mass and length, resulting in a Machian MOND formulation

\begin{equation}\label{eq3}
    g =\frac{c^2}{\left(\frac{M_u}{R_u}+\frac{M}{r}\right) \ \mu\left(\frac{M_u/{R_u}^2}{M/{r}^2}\right)} \frac{M}{r^2} \ \ or \ \ \frac{M}{r^2}= \frac{\left(\frac{M_u}{R_u}+\frac{M}{r}\right) }{c^2} \mu\left(\frac{M_u/{R_u}^2}{M/{r}^2}\right)g
\end{equation}

\begin{equation}\label{eq4}
    \mu\left(\frac{M_u/{R_u}^2}{M/{r}^2}\right) = \frac{1}{\sqrt{1+\frac{M_u/{R_u}^2}{M/r^2}}} \ \ or \ \ \mu\left(\frac{M_u/{R_u}^2}{M/{r}^2}\right) = \frac{1}{{1+\sqrt{\frac{M_u/{R_u}^2}{M/r^2}}}}
\end{equation}

in which for most cases where local potentials are small, $\left(\frac{M_u}{R_u}+\frac{M}{r}\right) \sim \left(\frac{M_u}{R_u}\right)$. The first Machian interpolating function is the best fit to Milgrom's MOND $\nu(a_0/a_N)$.

\newpage

\section{Discussion}\label{sec4}

Since only equivalent Machian substitutions for $G$ and $a_0$ are made yielding approximately the same numerical values (up to omitted numerical factors), and since $M$ and $r$ take the same meanings as in Milgrom's MOND, the resulting corrections (\ref{eq3}) and (\ref{eq4}) for galaxy rotation curves are equivalent to Milgrom's MOND at present epoch. For $\frac{M_u/{R_u}^2}{M/{r}^2}<<1$, $\mu\left(\frac{M_u/{R_u}^2}{M/{r}^2}\right)\rightarrow1$ and Newtonian laws are restored with $v=c \sqrt{({M/r})/({M_u/R_u})}$. For $\frac{M_u/{R_u}^2}{M/{r}^2} >> 1$,  $\mu\left(\frac{M_u/{R_u}^2}{M/{r}^2}\right)\rightarrow\sqrt{\frac{M/r^2}{M_u/{R_u}^2}}$, $v=c\sqrt[4]{M/M_u}$, and asymptotic rotational velocities are determined by the relative gravitationally active masses of the nearby and large scale mass distributions. 

This framework is free from dimensional constants and free parameters except for the speed of light, which takes the meaning of the speed of gravity and causality justifying the choice of $R_u$ as the radius of the observable universe containing mass $M_u$, which is the part of the universe causally affecting local dynamics.

The numerator $GM_u/{R_u}^2$ in (\ref{eq4}) prior to canceling of the gravitational constants with the denominator $a_N=GM/r^2$ does not represent a gravitational field intensity, since these must vanish for large scale: at any point in the universe there exists a surrounding observable volume sphere of approximately constant matter density whose net field intensity is zero at its center. Two Machian possibilities for term $GM_u/R_u^2$ are left. It arises as either the average acceleration of masses in the universe radially receding from the observer, or as a directionless sum of the inverse-squared distance contributions from all masses in the observable universe, times the gravitational constant. The former would imply that for the case of a non-expanding universe, MOND effects would vanish and that the decrease of inertia is due to such radial motion of the distant masses. The latter, which seems more straightforward, implies that $a_0=3GM_u/R_u^2$ from the integral over the spherical observable volume. Such quantity will be referred to as the universal field scale, which is a scalar measure of how strongly mass is concentrated around an evaluation point. In Machian theories, the gravitational constant and inertial mass are sourced by a scalar field composed of inverse-distance mass contributions $\Phi=\sum_{i}m_i/|r-r_i|$, commonly identified with the Newtonian potential. Within this perspective, MOND can be interpreted as a nonlinear additional contribution arising from the scalar inverse-squared distance mass terms $\phi=\sum_{i}m_i/|r-r_i|^2$. This introduces a stronger dependence of local dynamics on the global matter distribution than linear approaches with only Sciama’s relationship, making the framework more explicitly Machian.

The relationship pointed out by Milgrom between his MOND's acceleration scale constant $a_0$ and the cosmological constant or the Hubble parameter is interpreted as merely accidental due to the cosmological coincidence by which, at the present epoch, the energy densities associated with dark energy (the cosmological constant) and visible matter are of the same order of magnitude.

Although the origin of the Machian MOND function is not explained or derived here, it seems that it follows from Mach's principle's argument about which frame is acceleration defined with respect to. When local field intensities are above the universal field scale (at the cores of galaxies or within the solar system), the ‘main frame' acceleration is defined with respect to is the local one, and one can describe the dynamics of the system without referring to the scalar field intensity of the universe at large through the usual Newtonian laws (and Machian effects vanish). But, if the scalar sum of field intensities of the rest of the universe is greater than the field intensity of the local system (in the deep MOND regime, such as the flat velocity part in galaxies), it seems that acceleration is defined with respect to the background frame of the universe, and the field scale of the universe at large must be taken into account non-linearly together with that of the local frame, to explain the dynamics of the local system (in the case of Machian MOND, effectively through (\ref{eq4}) ). In between both regimes, the identification of the ‘main frame' is shared between the local and global frames, described approximately by the slope of the Machian MOND function.

Curiously, while a decrease of inertia occurs the smaller the Newtonian field is with respect to the universal field scale when $\frac{M_u/{R_u}^2}{M/r^2} >> 1$, which agrees with the Machian notion of a decrease of inertia when (gravitational) masses are removed from the neighborhood of the body, a decrease of $M_u$ implies a vanishing of MOND effects and an increase of inertia ‘Newtonizing' motion, which does not align with the Machian statement. However, by decreasing $M_u$ considering also Sciama's relationship for the gravitational constant, as introduced in the observable $v=c\sqrt[4]{M/M_u}$, the gravitational force increases (having the same observational consequence as considering Sciama's relationship alone when $\frac{M_u/{R_u}^2}{M/r^2} << 1$, which is the decrease of inertia).

For the non-physical case in which the observable universe were infinite ($M_u \rightarrow \infty$, $a_0=GM_u/R_u^2 \rightarrow \infty$), it corresponds to the deep MOND limit described pure MOND behavior in which the equations must be fully scale invariant. To achieve this, Milgrom sets $G\rightarrow 0$ keeping ${{\mathcal{A}}_0}\equiv G a_0$ constant, arguing that Newtonian gravity introduces a dimensional constant $G$ with a preferred scale that breaks scale invariance. But in Machian MOND, $M_u \rightarrow \infty$ directly implies $G=c^2/(M_u/R_u)\rightarrow 0$. This case would be the classical Newtonian cosmological model of an infinite, homogeneous, and isotropic universe, in which problems with the consistency of Newton's laws have for long been discussed. Relatively recently, it has been argued that the solution comes from the introduction of a sort of relativity of acceleration in the Newtonian laws \cite{Norton}.

The fact that Newtonian laws are recovered when the mass of the universe tends to zero ($M_u\rightarrow 0\,,a_0 \rightarrow 0$) fixing the gravitational constant, and that the MOND potential diverges at infinity, point towards the long-discussed problem of boundary conditions of the universe at large in GR, which has been linked to Mach's principle in the past (and Einstein's attempt to specify a divergence of the metric at infinity to satisfy the relativity of inertia).

But the mass of the universe can never be zero in the system under study, it can only be as small as the very same mass of the system itself. Decreasing the mass of the universe until the absence of any background ($M_u\rightarrow M, R_u\rightarrow r$), the function (\ref{eq4}) reduces to some simple number $n$ which can be set to unity with the appropriate choice of integers in the formulation of the Machian interpolating function and in Sciama's relationship, and velocities approach $v\rightarrow nc$. A simple way to obtain $n=1$ is considering Schrödinger's or Treder's definition of $G=c^2/(2M_u/R_u)$ and $a_0=3GM_u/{R_u}^2$ from the scalar sum of $1/r^2$ masses for a uniform matter density sphere of total mass $M_u$. This is already a Machian feature of Sciama's relationship, by which in absence of a background universe, rotation is undefined and tangential speed can take any value up to the limit of the speed of light. The proposed transformation respects this result as well as the invariance under global rescalings of mass, length, and time that Sciama's relationship introduces in the Newtonian regime. Machian MOND extends such invariance to the deep MOND regime.

Machian MOND satisfies several definitions of Mach's principle: Newton’s gravitational constant $G$ is a dynamical field, and consequently, the inertial mass of a body increases with the agglomeration of masses in its neighborhood. Moreover, there's an extra decrease of inertia a body is isolated from other masses (at low local gravitational field intensities) due to the interpolating function. The mass and size of the observable universe are always accounted for when considering non-inertial reference frames and local inertial frames are affected by the cosmic distribution of matter. Rotation is undefined up to the speed of light in the absence of the rest of the universe (or equivalently, a body in an empty universe has no inertia).

Milgrom himself favors modifying inertia (which is the true aim of Mach's principle), and not just gravity. As modified inertia, the inertial law in Machian MOND depends on the inertial mass of the body, the gravitationally active mass $M$ and the mass of the universe $M_u$, i.e., inertia depends on all mass distributions at play, in a Machian spirit. Once inertia is understood as having an origin in gravitation under the Machian reasoning, the issues of MOND as modified inertia by a kinetic term in function of the trajectory (such as non-localities due to the function requiring knowledge of the entire trajectory \cite{Milgrom1993}) no longer occur. Scalar matter fields, either inverse or inverse-squared distance ones, can be introduced in Newton's second law, without modifying the Poisson equation (and leaving gravity intact).

Milgrom's reluctance to explore a Machian version of MOND could be attributed to the widespread false claim that Mach's principle necessarily implies some sort of anisotropy of inertia. The anisotropy of inertial mass is strongly constrained by the Hughes-Drever experiments, and nearby nonuniform mass distributions, such as the Milky Way in experiments on Earth, do not show a preferred direction for inertia. Not only was it originally pointed out as a possible but not necessary consequence of Mach's principle \cite{Cocconi1958}, but the debate extended with Robert Dicke's claims about the unobservability of anisotropic inertial mass effects mass in relativistic theories and Treder's defense of scalar inertial mass in classical models, which prevents it from having a preferred direction.

MOND's interpolating function resembles the structure of a Lorentz-like transformation $\gamma=1/\sqrt{1-\beta}$. If a function depending on velocities, such as the Lorentz factor, arises from Lorentz invariance in SR, where accelerations are absolute, one would expect a function depending on accelerations or field intensities, such as Machian MOND, to arise when relativizing accelerations. The introduction of the constant speed of light from Maxwell's electromagnetism into relativity lead to Einstein's SR, and perhaps Milgrom's acceleration scale constant inspires a new relativity of accelerations. Lorentz invariance implies that the speed of light is constant in all inertial frames, and it seems that Machian MOND implies that the universal field scale at any point in the universe is always $>M_u/R_u^2$. It may be that Machian MOND arises from a new boost symmetry generalizing Lorentz invariance or from a transformation between accelerating frames, for instance, through the relativity of acceleration or the relativity of inertia, which is broken by the presence of the universe at large. Without a background universe, this symmetry would reduce to the Galilean or Lorentzian one.

Machian MOND, or a similar and equivalent approximate formulation, is considered to be a necessary non-linear feature that a phenomenological theory of modified inertia or modified gravity which incorporates Mach's principle reduces to as an approximation, in order to agree with galaxy rotation curves. This might arise by imposing Mach's principle (for instance, through the relativity of inertia of local objects, being determined by the sources of the gravitational field at the large scale) to a non-Machian non-linear theory such as GR (in which solutions to local dynamics do not depend on the mass or size of the observable universe). In fact, analogous interpolating functions in SR and GR have, as independent variables, dimensionless ratios of the form $v/c$ and $GM/Rc^2$, similar to Machian MOND. The form of the function (\ref{eq4}) is not unique and depends on the chosen Milgrom's MOND interpolating function for which the Machian substitutions are made. Although (\ref{eq4}) might be in conflict with solar system constraints, as it approaches $\mu(x) \rightarrow 1$ too slowly at the high acceleration regime, this is a problem of the slope of the function, which can be chosen to be different by another form for it. The function should arise naturally as an approximation, without the need of introducing it by hand (as introduced unnaturally in AQUAL or QUMOND), but the global $M_u$ and $R_u$ parameters governing local dynamics should appear in the complete FundaMOND theory, which is not the case in GR. So far, only linear and non-Lorentz invariant Machian theories of modified inertia with Sciama's relationship have been constructed, such as those of Treder and Schrödinger.

It is well known that, according to Bekenstein, a higher value for $a_0$ could resolve the tension between Milgrom's MOND and galaxy cluster dynamics (and that $a_0$ seems to grow with scale), where Newtonian accelerations take around that same value (provided that $a_0$ stays the same in galaxies for agreement with rotation curves). Machian MOND is a varying $a_0$ model with external mass distributions and could in principle help resolving this issue of Milgrom's MOND. This possibility will be discussed in a subsequent paper in preparation \cite{Santander}. It could also solve the cosmological issues of Milgrom's MOND \cite{Felten}, such as those arising from a fixed scale-dependence on $a_0$. As modified gravity, the equivalent variation of the Newtonian $G$ is both temporal due to the expanding universe and dependence on the mass and radius of the observable universe, and spatial since local potentials and local field intensities should also be included in Sciama's relationship and the Machian MOND function, respectively.

It would be interesting to compare Machian MOND with observational constraints of a varying effective $G$ and $a_0$ at the early universe, considering a cosmological model without most or all physical dark matter (assuming that MOND solves at least partly the need for dark matter), due to their dependence on the mass and radius of the observable universe, which have both varied over time. A similar study has been done for quantized inertia by showing the prediction of a specific increase in the galaxy rotation anomaly at higher redshifts \cite{McCulloch2017}, but quantized inertia only makes the $a_0$ of Milgrom's MOND dependent on $R_u$ through $a_o \sim c^2/R_u $. The formulation of quantized inertia is similar to the one proposed (without Sciama's relationship), but its interpretation is not based on Mach's principle. In fact, no Planck constant appears in the formulation, which suggests that its origin is not related to quantum phenomena, while the need to account for the universe at large directly points towards Mach's principle. More recently, it has been observed that there is no significant evolution in the TF law for redshifts up to $z \sim 2$ \cite{Stacy}, implying an almost constant $a_o$ in MOND. In contrast, Machian MOND shows a different dependence compared to quantized inertia: the Machian MOND function depends not only on $R_u$, but also on $M_u$. If the term $M_u/R_u^2$ is more constant than the simple dependence on $R_u$, this observation would not rule out Machian MOND (the observable radius was smaller in the past, but so was the mass contained in the volume defined by the observable radius). But as explained before, Machian MOND depends on many uncertainties, such as the Hubble tension, whether some physical dark matter is present in $M_u$, missing integers and Pi in the formulation, and relativistic effects of the universe as a whole.

\section*{Acknowledgments}
The authors acknowledge a similar but not exactly the same as that of equation (\ref{eq4}) to Juan D. Santander. We also thank Mordehai Milgrom for his warm and helpful feedback.

\section*{Declarations}

\text The authors have no competing interests to declare. This research did not receive any specific grant from funding agencies in the public, commercial, or not-for-profit sectors.

\bibliography{sn-bibliography}

@article{Milgrom1983a,
  author = "Mordehai Milgrom",
  title = "A modification of the Newtonian dynamics as a possible alternative to the hidden mass hypothesis",
  journal = "ApJ",  
  volume = "270",
  pages = "365--370", 
  year = "1983",
  doi = "10.1086/161130",
  publisher = "American Astronomical Society"
}

@misc{Deur2003,
      title = "Non-Abelian Effects in Gravitation", 
      author="A. Deur",
      year="2003",
      note = "Preprint at \url{https://arxiv.org/abs/astro-ph/0309474}",
}

@article{Brownstein_2006,
doi = {10.1086/498208},
url = {https://dx.doi.org/10.1086/498208},
year = {2006},
volume = {636},
number = {2},
pages = {721},
author = {J. R. Brownstein and J. W. Moffat},
title = "{Galaxy Rotation Curves without Nonbaryonic Dark Matter}",
journal = {ApJ},
}

@article{Verlinde2017,
	title="{Emergent Gravity and the Dark Universe}",
	author={Erik P. Verlinde},
	journal={SciPost Physics},
	volume={2},
	pages={016},
	year={2017},
	publisher={SciPost},
	doi={10.21468/SciPostPhys.2.3.016},
}

@article{Review,
title = "{Dark Matter: A Primer}",
author = "{Garrett K., Duda G.}",
doi = {https://doi.org/10.1155/2011/968283},
journal = {Adv. Astron.},
volume = {2011},
year = {2011},
}

@article{McCulloch2007,
title = "{Modelling the Pioneer anomaly as modified inertia}",
author = {McCulloch, M. E.},
doi = {https://doi.org/10.1111/j.1365-2966.2007.11433.x},
journal = {MNRAS},
number = {1},
pages = {338-342},
volume = {376},
year = {2007},
}

@article{Milgrom1993,
title = "{Dynamics with a Nonstandard Inertia-Acceleration Relation: An Alternative to Dark Matter in Galactic Systems}",
journal = {Ann. Phys.},
volume = {229},
number = {2},
pages = {384-415},
year = {1994},
doi = {https://doi.org/10.1006/aphy.1994.1012},
author = {M. Milgrom},
}

@article{Darabi2010,
title = "{A New Interpretation of MOND based on Mach Principle and Generalized Equivalence Principle}",
journal = {IJTP},
volume = {49},
pages = {1133–1139},
year = {2010},
doi = {https://doi.org/10.1007/s10773-010-0294-5},
author = {F. Darabi},
}

@article{Reissner1915,
title = "{Uber eine möglichkeit die gravitation als unmittelbarë Folge der relativität der trägheit abzuleiten}",
journal = {Phys. Z.},
volume = {16},
pages = {179–185},
year = {1915},
author = {H. Reissner},
}

@article{Schrodinger1925,
title = "{Die Erfüllbarkeit der Relativitätsforderung in der klassischen Mechanik}",
journal = {AdP},
volume = {382},
pages = {325-336},
year = {1925},
doi = {https://doi.org/10.1002/andp.19253821109},
author = {E. Schrödinger},
}

@article{Corda2016,
    author = "{Licata, I., Corda, C. and Benedetto, E.}",
    title = "A machian request for the equivalence principle in extended gravity and nongeodesic motion",
    journal = {Gravit. Cosmol.},
    volume = {22},
    pages = {48-53},
    year = "2016",
    doi = {https://doi.org/10.1134/S0202289316010102}
}

@book{Mach1881,
  title     = "Die Mechanik in ihrer Entwicklung",
  author    = "Ernst Mach",
  year      = 1881,
  publisher = "Brockhaus",
  address   = "Leipzig"
}

@article{Anderson2011,
title = {On the gravitodynamics of moving bodies},
journal = {Cent. Eur. J. Phys.},
volume = {9},
pages = {1151–1164},
year = {2011},
doi = {https://doi.org/10.2478/s11534-011-0030-7},
author = {Mol, Anderson W.},
}

@article{Sciama1953,
title = {On the origin of inertia},
journal = {MNRAS},
volume = {113},
pages = {34},
year = {1953},
author = {Sciama, D. W.},
doi = {http://dx.doi.org/10.1093/mnras/113.1.34}
}

@article{Whitrow1946,
title = "{The Mass of the Universe}",
journal = {Nature},
volume = {158},
pages = {165–166},
year = {1946},
doi = {https://doi.org/10.1038/158165b0},
author = {Whitrow, G.J.},
}

@article{WhitrowRandall1951,
title = "{Expanding World-models Characterized by a Dimensionless Invariant}",
journal = {MNRAS},
volume = {111},
pages = {455–467},
year = {1951},
doi = {https://doi.org/10.1093/mnras/111.5.455},
author = {Whitrow, G. J. and Randall, D. G.},
}

@article{Bondi1997,
title = "{The Lense-Thirring effect and Mach's principle}",
journal = {Phys. Lett. A},
volume = {228},
number = {3},
pages = {121-126},
year = {1997},
doi = {https://doi.org/10.1016/S0375-9601(97)00117-5},
author = {Hermann Bondi and Joseph Samuel},
}

@article{Dicke1961,
  title = "{Mach's Principle and a Relativistic Theory of Gravitation}",
  author = {Brans, C. and Dicke, R. H.},
  journal = {Phys. Rev.},
  volume = {124},
  issue = {3},
  pages = {925--935},
  year = {1961},
  doi = {10.1103/PhysRev.124.925},
}

@article{Gine2009,
title = "{On the origin of the inertia: The modified Newtonian dynamics theory}",
journal = {Chaos Solit. Fractals},
volume = {41},
number = {4},
pages = {1651-1660},
year = {2009},
doi = {https://doi.org/10.1016/j.chaos.2008.07.008},
author = {Jaume Gine},
}

@article{Gine2012,
title = "{The phenomenological version of modified Newtonian dynamics from the relativity principle of motion}",
author = {Jaume Gine},
doi = {https://dx.doi.org/10.1088/0031-8949/85/02/025011},
journal = {Phys. Scr.},
volume = {85},
number = {2},
pages = {025011},
year = {2012},
}

@article{Zwicky1933,
    author = "Zwicky, F.",
    title = "{Die Rotverschiebung von extragalaktischen Nebeln}",
    doi = "10.1007/s10714-008-0707-4",
    journal = "Helv. Phys. Acta",
    volume = "6",
    pages = "110--127",
    year = "1933"
}

@article{Rubin1983,
author = {Vera C. Rubin },
title = "{The Rotation of Spiral Galaxies}",
journal = {Science},
volume = {220},
number = {4604},
pages = {1339-1344},
year = {1983},
doi = {10.1126/science.220.4604.1339}

}

@article{Sanders2003,
    author = {Sanders, R. H.},
    title = "{Clusters of galaxies with modified Newtonian dynamics}",
    journal = {MNRAS},
    volume = {342},
    number = {3},
    pages = {901-908},
    year = {2003},
    doi = {10.1046/j.1365-8711.2003.06596.x},
}

@article{Sanders1998,
    author = {Sanders, R. H.},
    title = "{The Virial Discrepancy in Clusters of Galaxies in the Context of Modified Newtonian Dynamics}",
    journal = {ApJ},
    volume = {512},
    number = {1},
    year = {1998},
    doi = {10.1086/311865},
}

@article{Uzan2011,
    author = {Uzan, Jean-Philippe},
    title = "{Varying Constants, Gravitation and Cosmology}",
    journal = {Living Rev. Relativ.},
    volume = {14},
    number = {2},
    year = {2011},
    doi = {10.12942/lrr-2011-2},
}

@article{Felten,
       author = {{Felten}, J.~E.},
        title = "{Milgrom's revision of Newton's laws - Dynamical and cosmological consequences}",
      journal = {ApJ},
         year = 1984,
        doi = {10.1086/162569},
       volume = {286},
        pages = {3-6},
}

@misc{Unzicker2003,
  title = "{Galaxies as Rotating Buckets - a Hypothesis on the Gravitational Constant Based on Mach's Principle}",
  author = {Alexander Unzicker},
  eprint={gr-qc/0308087},
  archivePrefix={arXiv},
  primaryClass={gr-qc},
  year = {2003},
  note = "Preprint at \url{https://arxiv.org/abs/gr-qc/0308087}"
}

@article{Abreu2014,
title = "{Holographic considerations on a Machian Universe}",
journal = {Ann. Phys.},
volume = {351},
pages = {290-301},
year = {2014},
issn = {0003-4916},
doi = {https://doi.org/10.1016/j.aop.2014.09.004},
author = {Everton M.C. Abreu and Jorge {Ananias Neto}},
}

@book{Treder1972,
  title     = "Die Relativität der Trägheit",
  author    = "Hans-Jürgen Treder",
  year      = 1972,
  publisher = "De Gruyter",
  address   = "Berlin"
}

@book{Assis1999,
  title     = "Relational Mechanics",
  author    = "Andre Koch T. Assis",
  year      = 1999,
  publisher = "Apeiron",
  address   = "Montreal"
}

@article{Corda2024,
title = "{Equivalence Principle and Machian origin of extended gravity}",
journal = {Int. J. Mod. Phys. D},
year = {2024},
doi = {https://doi.org/10.1142/S0218271824410153},
author = {Benedetto, Elmo and Corda, Christian and Licata, Ignazio},
}

@article{Milgrom1983b,
  author = "Mordehai Milgrom",
  title = "A modification of the Newtonian dynamics—Implications for galaxies",
  journal = "ApJ",  
  volume = "270",
  pages = "371-383", 
  year = "1983",
  doi = "10.1086/161131",
  publisher = "American Astronomical Society"
}

@article{Milgrom1983c,
  author = "Mordehai Milgrom",
  title = "A Modification of the Newtonian Dynamics—Implications for Galaxy Systems",
  journal = "ApJ",  
  volume = "270",
  pages = "384-389", 
  year = "1983",
  doi = "10.1086/161132",
  publisher = "American Astronomical Society"
}

@article{McCulloch2017,
  author = "Michael E. McCulloch",
  title = "Galaxy rotations from quantised inertia and
visible matter only",
  journal = "Astrophys. Space Sci.",  
  volume = "362",
  number = "149", 
  year = "2017",
  doi = "https://doi.org/10.1007/s10509-017-3128-6",
}

@article{Stacy,
doi = {10.3847/1538-4357/ad834d},
year = {2024},
publisher = {The American Astronomical Society},
volume = {976},
number = {1},
pages = {13},
author = {McGaugh, Stacy S. and Schombert, James M. and Lelli, Federico and Franck, Jay},
title = {Accelerated Structure Formation: The Early Emergence of Massive Galaxies and Clusters of Galaxies},
journal = {ApJ},
}

@article{Brans1962,
  title = {Mach's Principle and the Locally Measured Gravitational Constant in General Relativity},
  author = {Brans, C. },
  journal = {Phys. Rev.},
  volume = {125},
  issue = {1},
  pages = {388--396},
  year = {1962},
  doi = {10.1103/PhysRev.125.388},
  url = {https://doi.org/10.1103/PhysRev.125.388}
}

@article{Einstein1918,
  title = {Prinzipielles zur allgemeinen Relativitätstheorie},
  author = {Einstein, Albert},
  journal = {AdP},
  volume = {55},
  pages = {241--244},
  year = {1918},
  doi = {10.1002/andp.19183600402},
  url = {https://doi.org/10.1002/andp.19183600402}
}

@article{Einstein1917,
  title = {Kosmologische Betrachtungen zur allgemeinen Relativitätstheorie},
  author = {Einstein, Albert},
  journal = {Sitzungsber. d. Königl. Preuss. Akad. d. Wiss. (Berlin)},
  pages = {142-152},
  year = {1917},
}

@book{Einstein1922,
  title     = "The Meaning of Relativity",
  author    = "Einstein, Albert",
  year      = 1922,
  publisher = "Princeton University Press",
  address   = "Princeton, NJ"
}

@article{Cocconi1958,
  title = {A search for anisotropy of inertia},
  author = {Cocconi, G. and Salpeter, E.},
  journal = {Il Nuovo Cimento},
  volume = {10},
  pages = {646--651},
  year = {1958},
  doi = {10.1007/BF02859800},
  url = {https://doi.org/10.1007/BF02859800}
}

@article{Norton,
 author = {John D. Norton},
 journal = {Philos. Sci.},
 number = {4},
 pages = {511--522},
 publisher = {[Cambridge University Press, The University of Chicago Press, Philosophy of Science Association]},
 title = {The Force of Newtonian Cosmology: Acceleration Is Relative},
 volume = {62},
 year = {1995},
 doi = {https://doi.org/10.1086/289883}
}

@article{Santander,
 author = "{M. Uruena Palomo and J. D. Santander}",
 journal = {In preparation},
 title = "{Machian MOND: a variable $a_0$ in galaxy clusters}",
 year = {2026}
}

\end{document}